\pdfoutput=1
\documentclass[aps,prl,twocolumn,groupedaddress]{revtex4-1}
\usepackage{calrsfs}
\usepackage{mathrsfs}
\usepackage{graphicx}
\usepackage{dcolumn}
\usepackage{bm}

\usepackage{hyperref}
\usepackage{xcolor}
\hypersetup{colorlinks=true,linkcolor=red}
\usepackage{color}
\usepackage{graphicx}
\usepackage{fancyhdr}
\usepackage{url}
\usepackage{amssymb,amsfonts,amsmath,amsbsy,bm,t1enc,lipsum,latexsym}

\newcommand{\eqn}[1]{Eq.(\ref{#1})}

\newcommand{\Fig}[1]{Figure \ref{#1}}

\newcommand{\fig}[1]{Fig. \ref{#1}}

\newcommand{\eqt}[1]{``#1''}

\begin{document}

\preprint{APS/123-QED}

\title{Raman induced soliton self-frequency shift in microresonator Kerr frequency combs}


\author{Maxim Karpov}
\affiliation{{\'E}cole Polytechnique F{\'e}d{\'e}rale de Lausanne (EPFL), CH-1015 Lausanne, Switzerland}

\author{Hairun Guo}
\affiliation{{\'E}cole Polytechnique F{\'e}d{\'e}rale de Lausanne (EPFL), CH-1015 Lausanne, Switzerland}

\author{Arne Kordts}
\affiliation{{\'E}cole Polytechnique F{\'e}d{\'e}rale de Lausanne (EPFL), CH-1015 Lausanne, Switzerland}

\author{Victor Brasch}
\affiliation{{\'E}cole Polytechnique F{\'e}d{\'e}rale de Lausanne (EPFL), CH-1015 Lausanne, Switzerland}

\author{Martin Pfeiffer}
\affiliation{{\'E}cole Polytechnique F{\'e}d{\'e}rale de Lausanne (EPFL), CH-1015 Lausanne, Switzerland}

\author{Michail Zervas}
\affiliation{{\'E}cole Polytechnique F{\'e}d{\'e}rale de Lausanne (EPFL), CH-1015 Lausanne, Switzerland}

\author{Michael Geiselmann}
\affiliation{{\'E}cole Polytechnique F{\'e}d{\'e}rale de Lausanne (EPFL), CH-1015 Lausanne, Switzerland}

\author{Tobias J. Kippenberg}
\email[E-mail: ]{tobias.kippenberg@epfl.ch}
\affiliation{{\'E}cole Polytechnique F{\'e}d{\'e}rale de Lausanne (EPFL), CH-1015 Lausanne, Switzerland}



\begin{abstract}
\label{sec_abstr}
  The formation of temporal dissipative solitons in continuous wave laser driven microresonators enables the generation of coherent, broadband and spectrally smooth optical frequency combs as well as femtosecond pulses with compact form factor. Here we report for the first time on the observation of a Raman-induced soliton self-frequency shift for a microresonator soliton. The Raman effect manifests itself in amorphous SiN microresonator based single soliton states by a spectrum that is hyperbolic secant in shape, but whose center is spectrally red-shifted (i.e. offset) from the continuous wave pump laser.
  The shift is theoretically described by the first order shock term of the material's Raman response, and we infer a Raman shock time of 20 fs for amorphous SiN. Moreover, we observe that the Raman induced frequency shift can lead to a cancellation or overcompensation of the soliton recoil caused by the formation of a (coherent) dispersive wave. The observations are in excellent agreement with numerical simulations based on the Lugiato-Lefever equation (LLE) with a Raman shock term. Our results contribute to the understanding of Kerr frequency combs in the soliton regime, enable to substantially improve the accuracy of modeling and are relevant to the fundamental timing jitter of microresonator solitons.

\end{abstract}


\maketitle

\label{sec_intro}
\emph{Introduction.} --- Microresonator based optical frequency combs (Kerr combs) \cite{delhaye2007comb,Kippenberg2011microcombs} enable optical frequency comb generation from a continuous wave (CW) laser, with repetition rates in the microwave domain (>10 GHz), and broad spectral bandwidth \cite{Okawachi2011octavespan,DelHaye2011octavecomb,moss2013cmos}.
Recently, a qualitatively new operation regime has been discovered \cite{herr2014soliton}, in which the parametrically generated comb seeds the formation of a temporal dissipative (cavity) soliton \cite{Lugiato1987structures,Wabnitz1993suppression, akhmediev2008dissipative}.
Such temporal dissipative solitons, first externally induced in fiber cavities \cite{leo2010temporal}, have been observed in Kerr frequency comb experiments using crystalline microresonators \cite{herr2014soliton,Herr2014avoidmodecross} and have recently also been generated in photonic chip based silicon nitride (SiN) integrated resonators \cite{moss2013cmos,Brasch2014Cherenkov}. Soliton based microresonator frequency combs have several attractive features, in particular being fully coherent, having smooth envelopes and giving access to femtosecond pulses. Indeed, the short duration of temporal solitons in crystalline microresonators have been used for external fiber broadening \cite{herr2014soliton,Jost2014link} and have allowed to achieve self-referencing (i.e. determination of the comb's carrier envelope frequency) enabling to count the cycles of light \cite{Jost2014link}. Moreover, taking advantage of dispersion engineering \cite{Coen2013modelling,Okawachi2011octavespan}, the presence of third (and higher) order dispersion allows (coherent) dispersive waves (DWs) to be generated \cite{Brasch2014Cherenkov} via the effect of soliton induced Cherenkov radiation \cite{akhmediev1995cherenkov}. This process has been used to create a photonic chip based (coherent) frequency comb in a SiN microresonator spanning 2/3 of an octave at electronically detectable mode spacing. However, while advances in theoretical simulations of the soliton regime have occured \cite{Hansson2014split,Coen2013modelling,Chembo2013LLE} a key underlying question is what other effects play a role in soliton formation and need to be included into the simulations.

In contrast to \eqt{Turing rolls} \cite{Coillet2013azimutalturing,godey2014stabilityLLE} dominating combs in the modulation instability (MI) regime, cavity solitons carry intense peak power and ultrashort duration such that in principle it could excite higher-order nonlinear effects such as the self-steepening effect and Raman scattering.
The latter usually exists and is broadband in nature in amorphous oxide and nitride materials such as silica and SiN, which relates to the vibrational material response of the cubic nonlinearity. Material Raman studies revealed typical active modes of e.g. mono-silicon bonds (${520 ~{\rm cm^{-1}}}$), silicon-oxygen bonds (${450 ~{\rm cm^{-1}}}$, broadband) \cite{lin2006raman} and silicon-nitrogen bonds (${400 ~{\rm cm^{-1}}}$ in amorphous ${\rm Si_3N_4}$) \cite{Wang2001RamanstudySiN}. In silica fibers, the interplay of the Raman effect with an ultrashort femtosecond pulse has been widely investigated \cite{Headley1996irsunified,Serkin2007ramanselfscat1} in the context of fiber-based supercontinuum generation and soliton propagation \cite{dudley2006supercontinuum}. The observed phenomena in soliton propagation in fibers include among others soliton self-frequency shift that continuously shifts the whole pulse spectrum to the red upon propagation \cite{Santhanam2003Ramanshift,Yulin2006irs}, soliton fission when a high-order soliton is split into fundamental solitons \cite{Yin2007solitonfission}, and  Raman induced pulse compressions in dispersive media \cite{Serkin2007ramanselfscat2}.
While well studied in fiber, to date however the role of the Raman effect in microresonator soliton formation has not been observed. Soliton formation in crystalline materials has not shown any evidence \cite{herr2014soliton,Herr2014avoidmodecross}. Recently, for amorphous silica- and SiN-based microresonators, the Raman effect was estimated by numerical simulations \cite{milian2015interplayRamanTOD,Zhang2014self-freqShift,Bao2014nonlinconversion}, and it was found that a comb spectral shift can occur when a Raman-nonlinearity assisted single temporal dissipative soliton is formed.
In this letter, we report for the first time to the authors' best knowledge the direct experimental observation of Raman induced soliton self-frequency shift in microresonator solitons.

\begin{figure}[t]
  \centering{
  \includegraphics[width = 1 \linewidth]{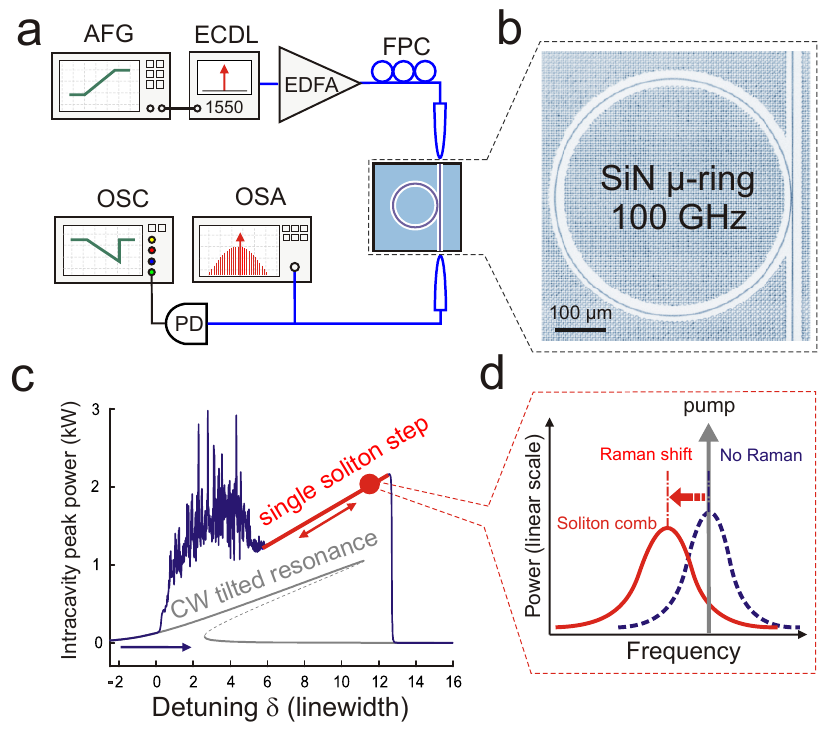}
  }
  \vspace{-8 mm}
  \caption{(a) Diagram of the experimental setup. AFG, arbitrary function generator; ECDL, external cavity diode laser; EDFA, erbium doped fiber amplifier; FPC, fiber polarization controller; OSA, optical spectrum analyzer; OSC, oscilloscope; PD, photodiode; (b) Optical microscope image of the SiN microresonator, with a ca. 100 GHz mode spacing. (c) Illustration of the pump laser detuning excitation scheme (reducing the pump frequency).
  The gray line shows the Kerr nonlinearity tilted cavity resonance profile with bistability referring to the CW, the blue line shows the trace of intracavity peak power with an increase of the cavity resonance--pump detuning ${\delta}$ (${\delta = f _{\rm 0}-f _{\rm p}}$ , where ${f _{\rm 0}}$ and ${f _{\rm p}}$ are the resonance and pump frequency), the red line shows the single soliton existence range.  (d) Illustration of the Raman induced spectral red-shift of the soliton frequency comb (in red line) compared to the case without Raman contributions (blue dashed).
  }
  \label{fig_diagram}
  \vspace{-5 mm}
\end{figure}

\emph{Experiments.} --- The study is based on SiN microresonators \cite{Levy2010compatible} in which recently temporal solitons were generated \cite{Brasch2014Cherenkov}. 
The first set of samples was fabricated by a newly developed \eqt{photonic damascene process} \cite{Pfeiffer2015Damascene}. 
Samples have a free spectral range (FSR) of 75--100 GHz on the fundamental ${\rm TE_{00}}$ mode family. Resonators have nominal SiN core height of ${0.9 ~{\rm \mu m}}$ and width ${1.65 ~{\rm \mu m}}$. The resonance linewidth ${\frac{\kappa}{2 \pi}}$ around 1550 nm is ${\sim 200 ~{\rm MHz}}$ that corresponds to a loaded ${Q}$-factor of ${\sim 10^{6}}$. The dispersion landscape is given by
${D_{\rm int}(\mu) = \omega _{\mu } - (\omega_0 + D_{1} \cdot \mu ) = \sum_{i>1} D_{i}\mu^{i}/{i!}}$, ${i\in \mathbb{N}}$ \cite{Delhaye2009disp}, where ${\mu}$ indicates relative mode index to the pumped mode (${\mu = 0}$), ${{\rm FSR} = \frac{D_{1}}{2\pi}}$, ${\omega_{\mu}}$ is the resonance frequency of the resonator, and ${D_i = \frac{\partial \omega_{\mu}}{\partial \mu }|_{\mu = 0}}$. At around 1550 nm, parameters ${\frac{D_2}{2 \pi} = 1 - 2  ~{\rm MHz}}$ (anomalous group velocity dispersion) and ${\frac{D_3}{2 \pi} = \mathcal{O}(1 ~{\rm kHz})}$ are characterized.



\begin{figure}[t!]
  \centering{
  \includegraphics[width = 1 \columnwidth]{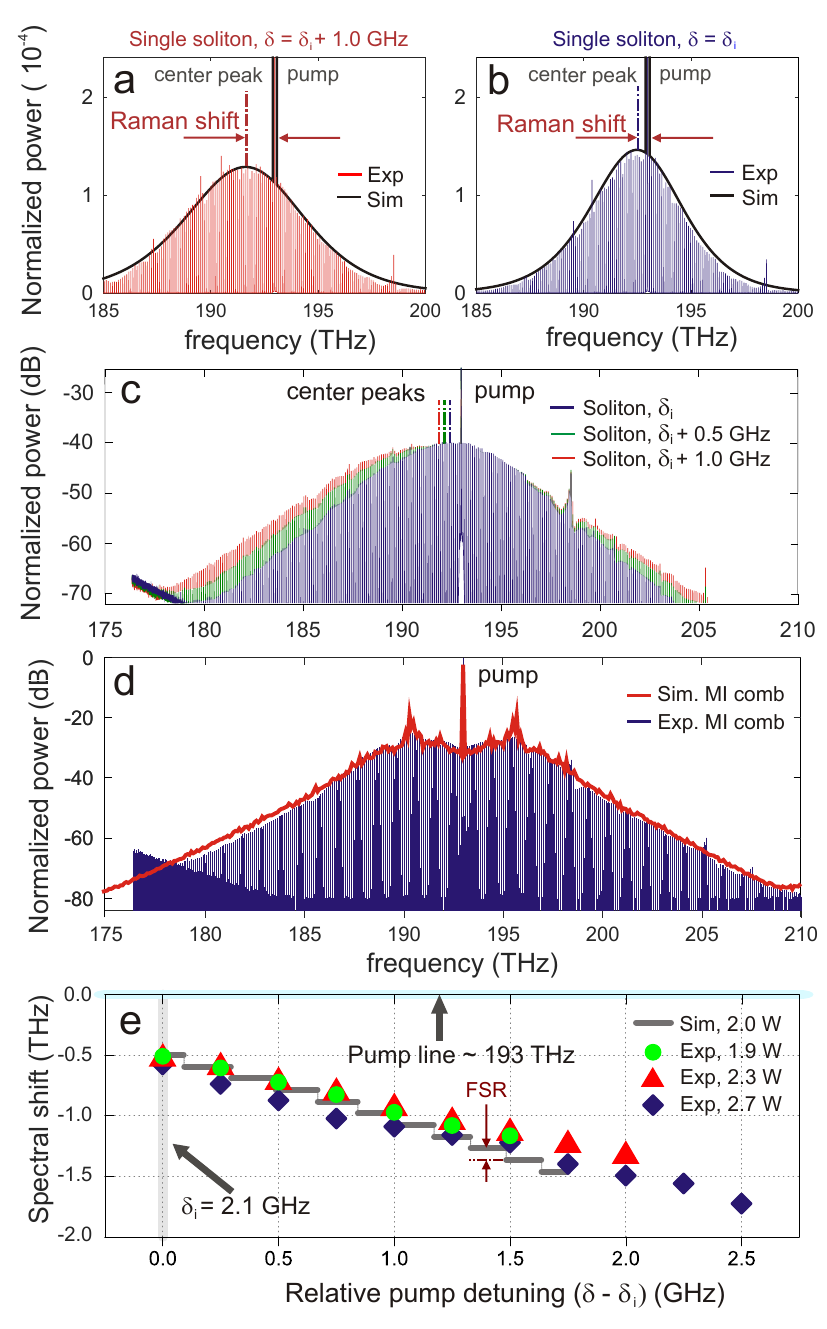}
  }
  \vspace{-8 mm}
  \caption{(a, b) Experimentally generated Kerr combs in the single temporal soliton regime in a 100 GHz mode spacing SiN microresonator, shown on a linear scale. The envelopes are from numerical simulations. The comb power (${y}$-axis) is normalized to the CW pump power. For soliton comb simulations, we used ${\frac{D_2}{2\pi} = 2 ~{\rm MHz}}$, ${\frac{D_3}{2\pi} = 4 ~{\rm kHz}}$, ${\frac{\kappa}{2\pi} = 350 ~{\rm MHz}}$, ${\delta _{\rm i} = 2.1 ~{\rm GHz}}$. (c) Measured soliton combs shown in logarithmic scale, red comb (same to (a)) has ${\delta  = \delta _{\rm i} + 1~{\rm GHz}}$, the green one corresponds to ${\delta  = \delta _{\rm i} + 0.5 ~{\rm GHz}}$, and the blue one (same to (b)) corresponds ${\delta  = \delta _{\rm i}}$. (d) Experimental measurement and simulation (the envelope) of a high noise MI comb, with the detuning ${\delta = 0.35 ~{\rm GHz}}$ outside the soliton existence range. (e) Variation of the comb spectral red-shift as a function of the detuning (${x}$-axis: ${\delta - \delta _{\rm i}}$), over the single soliton step at three pump powers: 1.9 W, 2.3 W and 2.7 W. Note that the trend from the simulation is discrete (gray line) as we mark the specific comb line of highest power (despite the strong pump line). The spacing corresponds to the FSR. ${\frac{D_1}{2\pi}}$}
  \label{fig_exp}
  \vspace{-5 mm}
\end{figure}

\fig{fig_diagram}(a) shows the experiment setup.
The CW pump is red-tuned to sweep over a resonance (at 1550 nm) with constant speed using an AFG (${1 ~{\rm nm/s}}$).
The pump power is 1--3 W on chip.
Usually, transitions to soliton states can be identified from the generated comb light \cite{herr2014soliton} that shows step-like pattern, indicating the formation of temporal dissipative solitons (e.g. \fig{fig_diagram}(c)).
Multiple solitons and single soliton are deterministically and repeatedly achievable by properly stopping the AFG-tuning.
Typical single soliton based frequency combs (soliton combs) in our SiN microresonators are shown in \fig{fig_exp}(a-c). The comb spectral envelope shows symmetric and smooth ${\mathrm{sech}^{2}}$-shape with a 3-dB bandwidth of ${\sim 5.5}$ THz, despite minor irregularities (e.g. the spike at 198.5 THz) caused by mode crossings \cite{Herr2014avoidmodecross}.
One striking difference to prior work in crystalline resonators \cite{herr2014soliton} is a global red-shift of the solitonic part of the comb spectrum, compared to the CW pump frequency, as shown in \fig{fig_exp}(a).

Once the single dissipative soliton is formed, one can further explore it's laser detuning dependence, by tuning the pump either to red or blue wavelengths, while the dissipative soliton still persists.
The range of laser detuning over which the soliton state exists (i.e. the \eqt{soliton step length}) usually corresponds to several GHz.

We then set the blue end of the single soliton step as the initial detuning (${\delta_i}$) and tune the pump wavelength over the entire laser detuning step length. With an increase of the detuning, the slowly evolving comb shows two apparent trends: first, the single soliton comb spectrum is broadened as the soliton duration is shortened and second, the global comb spectral red-shift  is increased. \Fig{fig_exp}(a, b) shows two comb spectra (in linear scale) at two positions within the single soliton step.
The comb red-shift in (a) is larger than that in (b) since the laser detuning ${\delta}$ is increased. \Fig{fig_exp}(c) shows the overall spectral broadening (in logarithmic scale) upon an increase of ${\delta }$.
For comparison, a high noise MI comb is also measured before the pump wavelength is tuned into the soliton existence range, see \fig{fig_exp}(d), in which the spectral shift is not observed (in agreement with the understanding that this regime does not produce short pulses inside the cavity).
Moreover, the spectral red-shift on single soliton combs is found to be changed by laser tuning within the single soliton step and such tunability is reversible by reversing the scan direction in contrast to the continuous red-shift in fiber optics \cite{Headley1996irsunified,Serkin2007ramanselfscat1}. 
The comb red-shift as a function of the detuning is shown in \fig{fig_exp}(e) for different pump powers. The shift ranges from 0.5 THz (${\sim 5 \times {\rm FSR}}$) to 1.75 THz (${\sim 17.5 \times {\rm FSR}}$) and exhibits almost linear dependence with the laser detuning in the single soliton existence range. 
The soliton step length is also increased with an increase of the pump power, shown in \fig{fig_exp}(e).

We attribute the comb spectral red-shift to the Raman induced soliton self-frequency shift \cite{Gordon1986theory,Headley1996irsunified,Serkin2007ramanselfscat1}, which in the studied SiN microresonator is well isolated and doesn't mix with higher-order dispersion effects, since the resonator produces weak third order dispersion and effectively suppresses the DW generation via soliton induced Cherenkov radiation \cite{akhmediev1995cherenkov}.
In addition, highly symmetric comb spectra imply weak self-steepening effects.

As a third order nonlinear effect, the Raman induced soliton self-frequency shift is in principle scaled by the soliton peak power, which matches the trend we observed in experiments.
 Analytically, the single temporal dissipative soliton consists of a CW background solution ${\Psi_{\rm CW}}$ and a superimposed temporal soliton ${\Psi_{\rm s} = \mathcal{C}\mathrm{sech}^{2}(\frac{t}{\Delta t})}$ where the soliton peak power is ${\mathcal{C} \propto \frac{\delta n^2 V_{\rm eff}}{c n_2}}$ (${n}$, ${n_2}$, ${V_{\rm eff}}$ are waveguide refractive index (RI), material nonlinear RI and effective mode volume, respectively), i.e. linearly proportional to the laser detuning ${\delta}$ \cite{herr2014soliton,Wabnitz1993suppression}. The soliton pulse duration, on the other hand, is ${\Delta t \propto \frac{D_2}{\delta \cdot D_1}}$ which is inverse proportional to the detuning \cite{herr2014soliton}. 
Therefore, benefitting from the increase of the soliton power and the decrease of the pulse duration upon a red-side pump wavelength detuning, the enhancement of the comb red-shift and the spectral broadening are expected.

\label{sec_sim}
\emph{Theory and numerical simulations} --- We numerically simulate the Kerr comb generation in SiN microresonators, based on LLE \cite{Lugiato1987structures,matsko2011mode,chembo2013spatiotemporal,herr2014soliton}, but include the Raman effect \cite{milian2015interplayRamanTOD,Zhang2014self-freqShift,Bao2014nonlinconversion}:
\begin{multline}
\frac{{\partial \tilde A(\mu ,t)}}{{\partial t}} = \sqrt {{\kappa _{{\rm{ex}}}}} {\left. {{s_{{\rm{in}}}}} \right|_{\mu  = 0}} - \left( {\frac{\kappa }{2} + i 2 \pi \delta } \right)\tilde A - i{D_{{\rm{int}}}}\tilde A \\
+ i\tilde gF\left[ {(1 - {f_{\rm{R}}})|A{|^2}A + {f_{\rm{R}}}({h_{\rm{R}}}(\phi ) \otimes |A{|^2})A} \right]
\label{eq_LLE}
\end{multline}
where ${|\tilde A(\mu )|^2}$ indicates the number of photons in the mode with index ${\mu }$ and ${t}$ is the propagation axis (i.e. the slow time frame).
\eqt{${\sim}$} marks the envelope written in frequency-like ${\mu }$ domain.
${A(\phi )}$ indicates the envelope in the phase ${\phi = \frac{2\pi \tau }{t_{\rm R}}}$ domain, where ${\tau }$ is the fast time frame and ${t_{\rm R} = \frac{2\pi }{D_1}}$ is the round-trip time.
Moreover, the input drive is ${s_{\rm in} = (P_{\rm in}/{\hbar \omega _0})^{1/2}}$ where ${P_{\rm in}}$ is the pump power. The cavity decay rate ${\kappa = \kappa _{\rm ex} + \kappa_{\rm 0} }$ is the sum of intrinsic decay rate ${\kappa_{\rm 0} }$ and coupling rate to the waveguide ${\kappa_{\rm ex}}$.
The nonlinear Kerr coupling is ${\tilde g(\mu ) = \hbar {\omega _0}\frac{{{\omega _\mu }c{n_2}}}{{n^2{V_{\rm eff}}(\mu )}}}$.
Here, ${\mu }$-dependent parameters (${n}$, ${n_2}$ and ${V_{\rm eff}}$) imply that the self-steepening effects can automatically be included in the model.
${F}$ indicates the Fourier transform,
${f_{\rm R}}$ is the Raman fraction and
${h_{\rm R}(\phi ) = {D_1}^{-1} h_{\rm R}(\tau )}$ is the Raman response scaled in ${\phi }$ domain where ${h_{\rm R}(\tau )}$ is in the physical (fast) time (${\tau }$) domain.
Therefore, the Raman spectrum has ${\tilde h_{\rm R}(\mu ) = \tilde h_{\rm R}(D_1^{-1} \omega )}$, where ${\omega }$ marks the physical frequency domain.
In the case that Raman active modes are sufficiently high in frequency and exceed the bandwidth of the soliton pulse, the resonant Raman effect plays a negligible role \cite{agrawal2013nonlinear}.
Then the Raman term can be simplified only containing the instantaneous response (direct current (DC) component) that contributes to the electronic Kerr effects, and the first-order component identified as the Raman shock term \cite{agrawal2013nonlinear}:
\begin{equation}
{h_{\rm R}} \otimes |A|^2 \approx |A|^2 - \frac{2\pi{\tau _{\rm R}}}{t_{\rm R}}\frac{\partial |A{|^2}}{\partial \phi }
\label{eq_shock}
\end{equation}
where ${\phi _{\rm R} = \frac{2\pi{\tau _{\rm R}}}{t_{\rm R}} = {i{{\tilde h}_R}^\prime (\mu = 0)}}$ and ${\tau _{\rm R}}$ is the material Raman shock time.
For the studied SiN microresonator samples, the fundamental single soliton is formed with a 3-dB bandwidth ${\sim 5.5}$ THz that is lower than the frequency of material Raman modes (${\sim 12}$ THz).

\begin{figure}[t]
  \centering{
  \includegraphics[width = 1 \columnwidth]{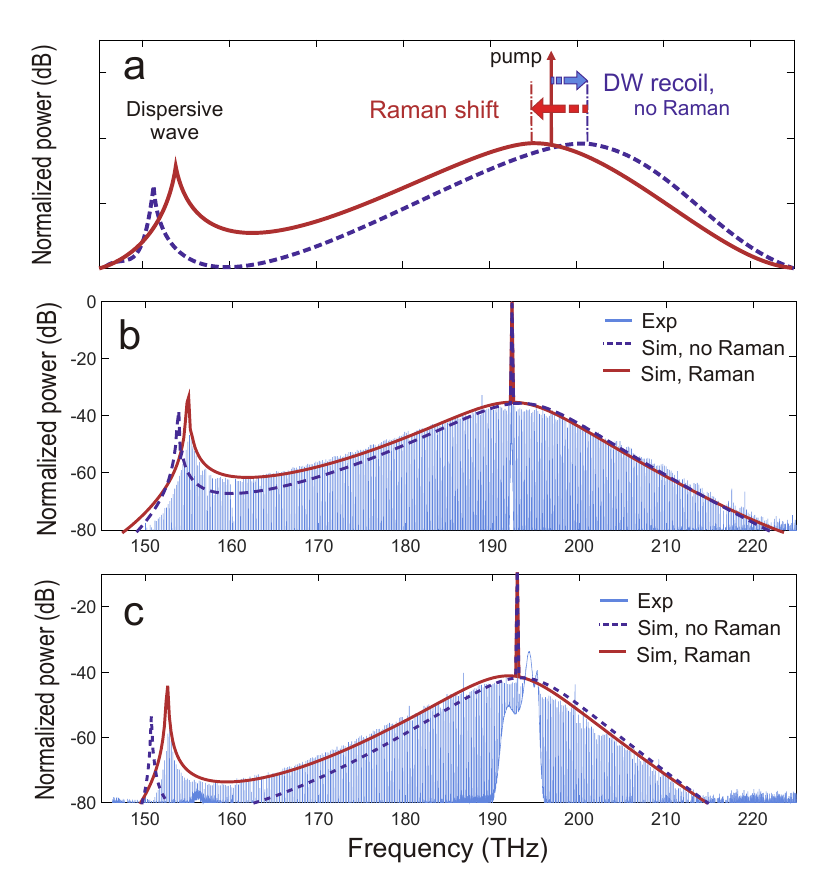}
  }
  \vspace{-8 mm}
  \caption{(a) Illustration of the cancellation effect between the Raman induced spectral red-shift and soliton blue-recoil through DW emissions on the long wavelength side. (b,c) Experimental generation and simulation of single soliton combs with coherent DWs in two different SiN microresonator geometries (both having a nominal SiN thickness of 0.8 ${\rm \mu m}$ and mode spacing of ${\sim }$190 GHz). Panel (b) shows the single soliton comb from the sample with a SiN core width of ${1.8 ~{\rm \mu m}}$ (sample taken from [12]), the comb 3-dB bandwidth is ${\sim 10}$ THz, ${D_{\rm int}}$ for the simulation contains ${\frac{D_2}{2 \pi} = 2.2 ~{\rm MHz}}$, ${\frac{D_3}{2 \pi} = 18 ~{\rm kHz}}$, ${\frac{D_4}{2 \pi} = -350 ~{\rm Hz}}$, and the pumped resonance is at 1560 nm employing 1 W of power 
  Panel (c) shows a soliton comb in a different sample with SiN core width of ${2.0 ~{\rm \mu m}}$ and the 3-dB bandwith is ${\sim 8}$ THz. Simulated dispersion parameters ${\frac{D_2}{2 \pi} = 3.2 ~{\rm MHz}}$, ${\frac{D_3}{2 \pi} = 26 ~{\rm kHz}}$, ${\frac{D_4}{2 \pi} = -340 ~{\rm Hz}}$. The pumped resonance is at 1554 nm and the power is 1 W. ${D_{\rm int}}$ and ${V_{\rm eff}}$ are calculated by finite element modelling using ${\texttt{COMSOL}{\textregistered}}$, while ${D_2}$ is also measured \cite{Delhaye2009disp}, showing close agreement to simulations. Blue dashed lines indicate comb envelopes without the Raman contribution.}
  \label{fig_DW}
  \vspace{-5 mm}
\end{figure}

Theoretically, the soliton self-frequency shift is fully described by the Raman shock term \cite{Gordon1986theory,Santhanam2003Ramanshift} that directly affects the pulse phase, thereby inducing a temporal phase shift that correspondingly red-shifts the pulse spectrum.
It is essential to delineate this process from the self-steepening effect that also introduces a shock term, but works directly on the pulse temporal amplitude.
An explicit self-steepening term could be recognized when the nonlinear coupling is written in ${\phi}$ domain, i.e. ${F^{-1}[{\tilde g}] = {\hat s} \cdot g_0}$, where ${g_0 = {\tilde g}(\mu = 0)}$.
${{\hat s} = 1-i \phi _{\rm s} \frac{\partial}{\partial \phi}}$ indicates the self-steepening operator, where the shock time (scaled in ${\phi }$ domain) is ${{\phi _{\rm{s}}} = \frac{{{D_1}}}{{{\omega _0}}} - \frac{{\Delta V}}{{{V_{{\rm{eff}}}(0)}}}}$, ${\Delta V}$ indicates the first-order variation of ${V_{\rm eff}(\mu)}$ (in ${\mu}$) around ${\mu = 0}$.

We expand \eqn{eq_LLE} into amplitude and phase dynamics of ${A(\phi)}$, using the ansatz ${A(\phi,t) = u(\phi,t)e^{i\varphi (\phi,t)}}$. Assuming the self-steepening as well as the Raman effect only make perturbations to the cavity single soliton that is the eigen-solution of the standard LLE (${D_{i>2} = 0}$, ${\phi _{\rm s} = 0}$, and ${f_{\rm R} = 0}$), we have following two relations:
\begin{equation}
\frac{{\partial \Delta u}}{{\partial t}} = 6{g_0}{\phi _{\rm{s}}}{u^2}\frac{{\partial u}}{{\partial \phi }}, ~ %
\frac{{\partial \Delta \varphi }}{{\partial t}} =  - 2{g_0}{f_{\rm{R}}}{\phi _{\rm{R}}}u\frac{{\partial u}}{{\partial \phi }}
\label{eq_amp}
\end{equation}
where ${\Delta u}$ and ${\Delta \phi}$ indicate the amplitude and phase perturbations.
It is clear that while the phase perturbation is induced by the Raman shock term ${\phi _{\rm R}}$, the amplitude perturbation is related to ${\phi _{\rm s}}$.
Self steepening thereby gives rise to shock front on the soliton pulse that distorts the spectrum and makes it asymmetric (opposite to the Raman case).
In experiments, a highly symmetric comb envelope (\fig{fig_exp}) actually implies weak self-steepening effects.
Reasons include that, first the comb bandwidth is small compared to the pump frequency, and second, the non-zero ${\Delta V}$ helps reducing the overall self-steepening effects.

\Fig{fig_exp} contains a set of numerical simulations which shows almost perfect agreement to the experiments.
We set the Raman fraction ${f_{\rm R} = 20 ~\%}$.
Importantly, the Raman shock time ${\tau _{\rm R}}$ in SiN is so far unknown and has not been reported to the authors' best knowledge.
However, close agremeent between theory and experiment and precise knowledge of experimental parameters allow us to determine the parameter to be ${2\pi \tau _{\rm R}\sim 20 ~{\rm fs}}$. Moreover, as independent verifications, this value is found to agree between different SiN microresonators, including those with significant different structural geometries.
By linearly tuning ${\delta }$, the comb evolution is simulated and transitions from primary comb, noisy MI comb to temporal dissipative soliton states are identified.
The trace of the intracavity peak power (cf. \fig{fig_diagram}(c)) shows the single soliton existence range ${2 \pi \delta = 6 - 11 {\kappa}}$.
\Fig{fig_exp}(a,b) show simulated comb envelopes in the soliton state, both with significant red-shifts.
In a high noise MI comb \cite{herr2014soliton} (\fig{fig_exp}(d)), however, the chaotic cavity light pattern is identified which is not as stable and intense as a cavity soliton, and the comb spectral red-shift is absent in agreement with experiment.
Simulations also verified the trend of the comb red-shift as a function of the detuning ${\delta}$, see \fig{fig_exp}(e).

It is interesting to note that the Raman shock time we extracted for SiN in this work (${2\pi \tau _{\rm R} \sim 20 ~{\rm fs}}$) is actually much smaller than that in silica, the latter being ${2\pi \tau _{\rm R} \approx 2\pi \times 89 ~{\rm fs}}$ \cite{lin2006raman}. A large shock time enables the observation of soliton self-frequency shift in silica-based fibers with short lengths (${\mathcal{O}(10)}$ cm) and intense pump powers (${\mathcal{O}(1)}$ kW) \cite{liu2001pulse}, while in SiN-based waveguides (not resonators), the Raman effect is not observed \cite{zhao2015visible,halir2012ultrabroadband}.
However, in microresonators, due to the cavity building (finesse ${\mathcal{F} = \frac{D_1}{\kappa} \approx \mathcal{O}(100 - 1000)}$), the light-material interaction length is effectively increased (to ${\mathcal{O}(1)}$ m) and meanwhile the pulse peak power is dramatically promoted (e.g. in \fig{fig_diagram}(c) the intracavity soliton peak power is up to ca. 3 kW under a CW pump of 2 W). Therefore, the weak Raman effect can be excited.
Moreover, the single soliton formation in a microresonator promises a persistent fundamental soliton pulse that could naturally suppress high-order Raman effects (e.g. the soliton fission) and leave the pure effect of soliton self-frequency shift, despite that the comb bandwidth may reach the frequency of Raman active modes.
To this point, the Raman effect actually assists the formation of the soliton comb, provided that the Raman gain doesn't exceed the cavity loss (see a recent careful study in \cite{milian2015interplayRamanTOD}). Therefore, even though the shock time is much shorter than that in silica glass, we could still observe the Raman induced soliton self-frequency shift in SiN microresonators.

\label{sec_DW}
\emph{Cancellation of the soliton spectral recoil by the Raman shift.} --- We next investigate how the Raman effect influences cavity soliton with the presence of higher-order dispersion effects.
For this we use two other sets of SiN microresonators with two different structural geometries that allow shorter soliton pulses and enable coherent DW generation, as recently demonstrated in \cite{Brasch2014Cherenkov}.
Third order dispersion ${\frac{D_3}{2\pi} = \mathcal{O} (10) ~\rm kHz}$ is characterized in the waveguide, on the fundamental ${\rm TM_{00}}$ mode.
A zero dispersion wavelength is estimated at ca. 1700 nm. Experimentally pumped at 1550 nm a single soliton comb is generated in both samples, and more importantly, a coherent DW is identified at around 195 THz (cf. \cite{Brasch2014Cherenkov} for more experimental details).
Since a coherent DW is spectrally situated in the normal dispersion regime, the overall comb bandwidth is significantly extended to ${\sim 75}$ THz.
The comparison of the two spectra in \fig{fig_DW}(b) and \fig{fig_DW}(c) shows the influence of dispersion on the position of the DW frequency. 

It is well understood that the emission of a DW will lead to the soliton spectral recoil \cite{akhmediev1995cherenkov, erkintalo2012cascaded}, such that the comb profile around the CW pump frequency will be shifted away from the DW, as illustrated schematically in \fig{fig_DW}(a). However, in above mentioned SiN microresonators, the blue-recoil (resulting from the DW being on the red side) was surprisingly \emph{not} observed \cite{Brasch2014Cherenkov}, see \fig{fig_DW}(b), and even a red-shift (\fig{fig_DW}(c)) of the comb profile is seen. The explanation for this unusual observation is that the Raman induced soliton self-frequency shift cancels and even overcompensates the soliton blue-recoil. Indeed, a similar effect has been observed in the case of soliton propagation in optical fibers \cite{skryabin2003soliton}. We compare the two spectra to numerical simulations including both the Raman shock term and the full dispersion landscape ${D_{\rm int}}$, and we keep the previously determined value of the Raman shock time (20 fs). The simulations are in good agreement and in particular, reproduce the experimentally observed cancellation and overcompensation of the soliton spectral recoil by the Raman induced red-shift.

\label{sec_con}
\emph{Conclusions.} --- %
In summary, we experimentally observed and studied for the first time the Raman-induced soliton self-frequency shift in SiN microresonators.
Concerning the implications of our work, the results enable to increase substantially accuracy and understanding in numerical modeling of soliton based frequency combs in the soliton regime for SiN in particular, and amorphous microresonators in general. The observations should apply to other classes of amorphous microresonators, such as aluminum nitride or silica based. Likewise  the observation of Raman induced red-shifted soliton spectra has implications for understanding the fundamental noise properties such as to the timing jitter of the emitted solitons (and thus phase noise in the generated microwave beatnotes), as the red-shift implies an energy loss to the phonon and results in additional noise as governed by the fluctuation dissipation theorem \cite{callen1951irreversibility}.


\label{sec_ack}
\emph{Acknowledgements.} --- This work was supported by the DARPA PULSE Program, an European Space Agency TRP and the Swiss National Science Foundation. M.K. acknowledges the support from the Marie Curie Initial Training Network FACT, M.G. the support from the Hassler foundation and the Marie Curie co-Fund program at EPFL. All samples were fabricated at the center for MicroNanotechnology CMi at EPFL.




%

\end{document}